\documentclass[conference]{IEEEtran}

\usepackage[T1]{fontenc}
\usepackage{amsmath}
\usepackage[cmintegrals]{newtxmath}
\usepackage{graphicx}
\usepackage{xcolor}
\usepackage{tabu}
\usepackage{subfigure}
\usepackage{epstopdf}
\usepackage[sort]{cite}
\usepackage{ifthen}
\newboolean{bombastic}
\setboolean{bombastic}{false}

\graphicspath{ {Image/} }

\newtheorem{thm}{\textcolor{black}{Theorem}}
\newtheorem{lem}{\textcolor{black}{Lemma}}

\def \myref[#1]{\ifthenelse {\boolean{bombastic}}{\ref{#1} (#1)}{\ref{#1}}}

\begin{document}

\title{Maximizing Ergodic Throughput in Wireless Powered Communication Networks}

\author{\IEEEauthorblockN{Arman~Ahmadian, and Hyuncheol Park}
\IEEEauthorblockA{School of Electrical Engineering, KAIST\\
Daejeon, Republic of Korea\\
Email: \{a.ahmadian, hcpark\}@kaist.ac.kr}}

\maketitle

\begin{abstract}
This paper considers a single-antenna \emph{wireless-powered communication networ}k (WPCN) over a flat-fading channel.
We show that, \textcolor{black}{by using our probabilistic harvest-and-transmit (PHAT) strategy, which requires the knowledge of instantaneous full \emph{channel state information} (CSI) and fading probability distribution}, the ergodic throughput of this system may be greatly increased relative to that achieved by the \emph{harvest-then-transmit} (HTT) protocol.
To do so, instead of dividing every frame to the uplink (UL) and downlink (DL), the channel is allocated to the UL \emph{wireless information transmission} (WIT) and DL \emph{wireless power transfer} (WPT) based on the estimated channel power gain.
In other words, based on the fading probability distribution, we will derive some thresholds that determine the association of a frame to the DL WPT or UL WIT. More specifically, if the channel gain falls below or goes over these thresholds, the channel will be allocated to WPT or WIT.
%More specifically, when the channel power gain is lower than a threshold, the channel is allocated to the UL WIT and when the channel power gain is higher than the threshold, the channel is allocated to the DL WPT.
%Such a scheme results in two thresholds.
%If the channel power gain is either higher than the higher threshold or lower than the lower threshold, the channel is allocated to WPT and if the channel power gain is between the two thresholds, it is allocated to WIT.
Simulation results verify the performance of our proposed scheme.
\end{abstract}


\section{Introduction}
\ifthenelse {\boolean{bombastic}}
{
\color{olive}
\subsection{Motivation}
\color{black}
}{}
Recently \emph{wireless power transfer} (WPT) has been widely investigated.
Specifically, the use of WPT combined with \emph{wireless information transmission} (WIT) has found many applications such as wireless sensor networks (WSN), radio-frequency identification (RFID) systems and internet of things (IOT) networks \cite{07462480}.
WPT transfers energy wirelessly to supply the energy of such devices and hence enables them to function without batteries and thus truly wirelessly.
This area is called \emph{wireless powered communication} (WPC) and is further divided into two categories.
In the first category, the energy and information are transferred in the same direction and via the same RF signal.
This scheme is named \emph{simultaneous wireless information and power transfer} (SWIPT) \cite{06489506}.
The harvested energy is used to power up the device for wireless information reception.
In the second category, termed \emph{wireless powered communication network} (WPCN)\cite{06678102}, the energy and information are transferred in different directions.
Usually, the energy \emph{access point} (AP) transmits wireless energy to the \emph{wireless device} (WD).
The WD, on the other hand, uses the harvested energy to send some information to a data AP.
For instance, a medical sensor implanted in body may be powered wirelessly.
Such a sensor uses the harvested power to transmit medical information to a receiver outside.
In this case, the function of WPT is to cater to the sensor and its transmitter to operate.
Moreover, there are a number of works that propose full-duplex communication between the energy harvester and the energy transmitter \cite{07937876}.

\ifthenelse {\boolean{bombastic}}
{
\color{olive}
\subsection{Literature Review}
\color{black}
}{}
There exist a variety of different topologies and assumptions in WCPN.
\ifthenelse {\boolean{bombastic}}
{
\subsubsection{Multiple Antennas}
}{}In \cite{06489506, 07937876, 07009979, 06878442, 06623072, 16638567} the AP uses multiple-antenna techniques to focus the electromagnetic (EM) wave into a narrow beam while in \cite{06678102, 07393872} single (double)-antenna WDs and APs are considered\footnote{Double-antenna WDs are only proposed to separate the energy harvesting hardware from that of information transmission and so the analysis of such a WD is essentially identical to that of single-antenna WDs.}.
\ifthenelse {\boolean{bombastic}}
{
\subsubsection{Multi-user}
}
{}Multi-user design is considered in \cite{06878442, 07009979} while a single user case is studied in \cite{07937876, 07393872, 06623072, 16638567}.
\ifthenelse {\boolean{bombastic}}
{
\subsubsection{Fairness}
}{}
However, multiple WDs raise the fairness question which is considered in \cite{07009979}.
\ifthenelse {\boolean{bombastic}}
{
\subsubsection{CSI}
}
{}In \cite{06878442, 07393872, 07009979, 16638567} imperfect \emph{channel state information} (CSI), and in \cite{06678102, 06623072} full channel CSI is assumed to be available.
Moreover, \cite{07937876} studies both conditions.
Yet, to the authors' knowledge, the CSI probability distribution has never been exploited.

\ifthenelse {\boolean{bombastic}}
{
\subsubsection{Maximization criterion}
}{}
The objective parameter is also different in these works.
In \cite{06623072}, for example, the energy efficiency of the uplink (UL) throughput and in \cite{06678102, 07937876, 07009979, 07393872, 06878442, 16638567} the UL throughput itself has been maximized.
Yet, the throughput in a wireless channel is subject to variations in the channel gain.
The average, or the expected throughput is termed \textit{ergodic throughput} \cite{Goldsmith} which has been considered in \cite{07937876, 07393872}.
Yet, in \cite{07937876} instantaneous full channel CSI is unavailable and in \cite{07393872} the problem has been restricted to an OFDM-based system.
In addition, in neither of these two works the fading distribution has been utilized.
In \cite{07393872} for example, the maximization has been done for a fixed set of given channel power gains and the assumed parameter is average transmit power.
Both of these assumptions are different in this paper.

\ifthenelse {\boolean{bombastic}}
{
\color{olive}
\subsection{Challenges}
\color{black}
}{}
Nonetheless, for the most basic form of a WCPN, that is a single WD and a \textit{hybrid access point} (HAP)\footnote{\color{black}In a WPCN, the data and energy access points may be collocated, in which case the AP is called a hybrid access point.\color{black}} both with single antennas, there are a number of challenges to solve.
The most important problem is how to divide the communication channel into downlink (DL) WPT and UL WIT.
This is an important issue because as more resources are allocated to the DL, the WD receives more power and hence has enough energy to transmit back its information to the AP.
However, as more and more resources are allocated to the DL, there may not be sufficient resources left at the UL and so the UL data rate may decrease.

So far, the best-known method to share the communication channel resources between DL and UL has been the \emph{harvest-then-transmit} (HTT) protocol \cite{06678102, 16638567, 06878442}.
According this method, based on the estimated channel power gain, a ratio of every frame is allocated to the DL WPT and the rest to the UL WIT, hence, allowing the WD to first harvest energy and then transmit information using the harvested energy in the first phase.
This method mandates that the WD both harvest as well as transmit during each frame.
However, as mentioned, the channel power gain in a wireless channel varies probabilistically.
If the channel power gain probability distribution is known, \textcolor{black}{using our method, called the probabilistic harvest-and-transmit (PHAT),} this information may be exploited to yield a higher ergodic throughput in fading channels.
The key factor to exploit here is to use the channel for DL WPT rather than UL WIT when the channel condition favors DL WPT and vice versa.
The question of when the channel is more appropriate to WPT rather than WIT is answered by solving an optimization problem that we will subsequently derive.

\ifthenelse {\boolean{bombastic}}
{
\color{olive}
\subsection{Outline}
\color{black}
}{}
In section two of the paper, the system model of the problem is described for \textcolor{black}{both the HTT protocol as well as the PHAT scheme.}
In section three, the HTT protocol and the proposed scheme's optimization problems is presented.
However, the proposed scheme's optimization problem is not a convex one.
Therefore we also present a simple convex subclass of the problem.
In the final section, it is shown through simulation that the proposed scheme can lead to a significant increase in the throughput of WPCN.
The paper ends with a conclusion and discussion in section five.

\section{System Model}
\begin{figure}[t]
\includegraphics[trim=230 250 450 190,clip,scale=0.8]{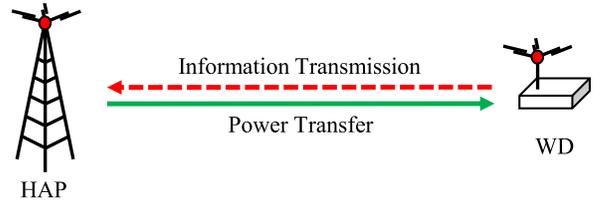}
\centering
\caption{Schematic of our single-user wireless communication network.}
\label{fig:schematic}
\end{figure}

We consider a WD and a HAP both equipped with a single antenna.
The HAP sends wireless power to the WD in the DL, and the WD harvests this power and use it to send back some information to the access point.
The setup is shown in Fig \ref{fig:schematic}.
It is assumed that the HAP is connected to some constant power source with a maximum transmit power of $p_d$ but the WD has no power source other than the wireless power received from the HAP.
The UL transmitter power of the WD for frame \emph{i} is $p_{u,i}$ and can vary from frame to frame.
We also assume that the WD has a lossless capacitor or battery with an infinite capacity to store the harvested energy and use it for UL transmission.
We consider a block flat fading model for the UL and DL channels where the channel power gain remains constant during a single frame (block) but can vary from one frame to the other.
We define the \emph{normalized channel power gain} for frame $i$ as
\begin{equation*}
g_i={\left|{\tilde{g}_i}\right|}^2/\bar{g}\,,~~ i\in\mathbb{N}\,,
\end{equation*}
where $\bar{g}=E\left\{{\left|{\tilde{g}_i}\right|}^2\right\}$ is the average (expected) channel power gain and $\tilde{g}_i$ is the channel gain for frame $i$.
We assume $g_i$ is estimated perfectly and is known at both the HAP and the WD.

Let $f_G\left(g\right)$ denote the probability distribution function of the normalized channel power gain.
For simplicity of analysis, we assume the fading distribution of the channel is Rayleigh so that the normalized channel power gain distribution becomes exponential
\begin{equation*}
f_G(g_i) = e^{-g_i}\,,~~ i\in\mathbb{N}\,.
\end{equation*}
%
% Base on this definition, the expected SNR at the HAP can be calculated as $\bar{\gamma} = p_d\bar{g}/\sigma^2$.

We assume frame-based transmission with unity bandwidth and frame length.
In the rest of this section we explain  the system model specific to the two transmission strategies HTT and PHAT.

\subsection{Harvest-then-transmit protocol}
Based on the HTT protocol, in the $i$-$th$ frame, first DL WPT will occur for $\tau_i T$ seconds, allowing the WD to store $\mathcal{E}_{h,i}$ Joules of energy into the capacitor.
The WD then uses $\mathcal{E}_{c,i}$ Joules of energy to transmit information back to the HAP during the next phase of length $(1-\tau_i)T$ seconds.
Here, $\tau_i \in [0~~1]$ is the optimization variable that, as we will later discover, depends on the value of the normalized channel power gain.
\color{black} The objective is to achieve the maximum UL data rate at each frame.
\color{black}
In the rest of this paper, for the sake of brevity we will omit the frame subscript $i$ when describing the HTT protocol.

\subsection{Probabilistic-harvest-and-transmit method}
In our proposed method, rather than dividing every frame to UL and DL, a single frame is either associated with DL WPT or UL WIT.
The criterion based on which we make such a decision is the value of normalized channel power gain.
More specifically, if normalized channel power gain $g_i$ belongs to a \emph{WPT normalized channel power gain} set $\mathcal{P}\in\mathcal{R}_+$, then WPT will occur in the $i$-$th$ frame, whereas WIT will happen in this frame if the normalized channel power gain $g_i$ belongs to a \emph{WIT normalized channel power gain set} $\mathcal{I}\in\mathcal{R}_+$ where $\mathcal{R}_+$ is the set of non-negative real numbers.
The idea is that such a partitioning can potentially allocate the channel to WPT when the channel is more suitable for power transfer rather than information transmission and vice versa and hence maximize the ergodic UL data rate.

\section{Solutions}
In this section, first the mathematical formulation of the HTT protocol along with a closed-form solution is presented.
Then, a formal formulation of the PHAT method is given.
It is shown that this general formulation is not convex.
As a result, a simple sub-class of the problem is presented and shown to be quasi-convex.

\subsection{Harvest-then-transmit protocol}
As mentioned, in every frame, the WD both harvests energy and transmits information.
We will study the DL and UL constraints as well as the solution here.
\begin{itemize}
\item
\emph{Downlink}:
The harvested energy in each frame is
\begin{equation}
\mathcal{E}_{h}=\tau p_d \|\tilde{g}\|^2=\tau p_d \bar{g} g\,.
\end{equation}
\item
\emph{Uplink}:
The consumed energy in frame $i$ is
\begin{equation}
\mathcal{E}_{c} = \left(1-\tau \right)p_{u}\,,
\end{equation}
The UL data rate for frame $i$ is
\begin{equation}
r_\mathrm{HTT}(g) = (1-\tau)\log_2\left(1+ p_u \frac{g\bar{g}}{\sigma^2} \right)\,,
\end{equation}
where $\sigma^2$ is the noise variance.
In a flat fading channel, the ergodic throughput is given by
\begin{equation}
\overline{r}_\mathrm{HTT}=\int_0^{\infty}{r_\mathrm{HTT}(g)f_G(g)}dg\,.
\label{eq:Ergodic}
\end{equation}

\item
\emph{Constraint}:
The amount of consumed power $\mathcal{E}_c$ should be smaller or equal to the amount of harvested power $\mathcal{E}_h$ in each frame.
We assume the other circuitry in the WD receiver consume negligible power, and as a result we can presume they are equal.
In other words
$\mathcal{E}_c = \mathcal{E}_h$
which gives
\begin{equation*}
p_u = \frac{\tau }{1-\tau }p_d g \bar{g}\,.
\end{equation*}

\end{itemize}

Here, the question is how we should divide the time between UL WIT and DL WPT to maximize the UL data rate?
Therefore, the maximization problem for each frame is
\begin{subequations}
\begin{align}
& \underset{\tau}{\text{maximize}} && r_\mathrm{HTT}(g)=(1-\tau) \log_2 \left(1 + \gamma \frac{\tau}{1-\tau}\right)\,,  &&\label{eq:HTT}\\ 
& \text{subject to} && 0 \le \tau \le 1\,,
\end{align}
\label{pr:HTT}
\end{subequations}
where $\gamma = p_d\bar{g}^2g^2/\sigma^2$ is the instantaneous SNR at each frame.
As can be seen in (\ref{eq:HTT}) by increasing $\tau$ the harvested energy and hence the SNR increases.
Doing so, however, decreases WIT time, the former effect increasing the data rate and the latter decreasing it.
In summary, there is a trade-off in choosing $\tau$ and there exists an optimal $\tau$ maximizing the UL data rate.
The convexity of this problem is shown in \cite{06678102}.
\ifthenelse {\boolean{bombastic}}
{
\color{olive}
In the following lemma we provide a simple closed form solution for this problem.
}{}
\ifthenelse {\boolean{bombastic}}
{
\begin{lem}
\color{olive}
Optimal Value of the HTT method is given by
\begin{equation}
\tau = \frac{\gamma - 1 - W_0\left(\frac{\gamma-1}{e}\right)}{\left(W_0\left(\frac{\gamma-1}{e}\right)+1\right)(\gamma  - 1)}
\label{eq:THH_Optimum_1}
\end{equation}
where $W_0\left(\cdot\right)$ is the principal branch of the Lambert-W function.

\emph{Proof}: Taking the derivative of (\ref{eq:HTT}), and setting the result to zero we get
\begin{equation}
\frac{\gamma+\gamma\frac{\tau }{1-\tau }}{1+\gamma \frac{\tau }{1-\tau }}={\ln \left(1+\gamma \frac{\tau }{1-\tau }\right)\ }
\end{equation}
or
\begin{equation}
\frac{1}{1+\gamma \frac{\tau }{1-\tau }}e^{\frac{\gamma + \gamma \frac{\tau }{1-\tau }}{1+\gamma \frac{\tau }{1-\tau }}}=1
\end{equation}
which can be written as
\begin{equation}
\frac{ \gamma - 1}{1+ \gamma \frac{\tau }{1-\tau }}e^{\frac{ \gamma - 1}{1+ \gamma \frac{\tau }{1-\tau }}}=\frac{ \gamma - 1}{e}
\end{equation}
the solution to which is
\begin{equation}
\frac{\tau }{1-\tau } = \frac{\gamma - 1}{\gamma W_0\left(\frac{\gamma-1}{e}\right)}-\frac{1}{\gamma}
\label{eq:THH_Optimum_2}
\end{equation}
which can be written as (\ref{eq:THH_Optimum_1}).
\hfill $\blacksquare$
\end{lem}
\color{black}
}

\subsection{Probabilistic-harvest-and-transmit method (general formulation)}

In what follows, we first analyze the DL and UL channels and the constraints.
After that we present the optimization problem.

\begin{itemize}
\item
\emph{Downlink}:
The expected amount of harvested energy in the DL WPT is given by

\begin{equation*}
\mathcal{E}_h = p_d\bar{g}  \int_{\mathcal{P}} g f_G\left(g\right) dg\,.
\end{equation*}

\item
\emph{Uplink}:
The expected data rate (ergodic throughput) in the UL is given by
\begin{equation*}
\bar{r}_\mathrm{PHAT} = \int_{\mathcal{I}}\log_2 \left(1+ p_u\frac{g\bar{g}}{\sigma^2} \right)\ f_G\left(g\right)dg\,,
\end{equation*}
The expected amount of consumed energy is given by
\begin{equation*}
\mathcal{E}_c =\int_{\mathcal{I}} p_u f_G\left(g\right) dg\,,
\end{equation*}
where $p_u$ can, in general, be a function of $g$.

\item
\emph{Constraints}:
The WD cannot simultaneously harvest energy and transmit information.
Hence, the WIT and WPT normalized channel power gain sets should not overlap.
Specifically,
\begin{equation*}
\mathcal{I} \cap  \mathcal{P} = \varnothing\,.
\end{equation*}
Nevertheless, to maximize the system throughput, these sets should comprise the whole normalized channel power gain set $\mathcal{R}_+$.
In other words
\begin{equation*}
\mathcal{I} \cup \mathcal{P} = \mathcal{R}_+\,.
\end{equation*}
Furthermore, similar to the assumptions made in the last subsection, the amount of expected consumed power $\mathcal{E}_c$ should be smaller than or equal to the amount of expected harvested power $\mathcal{E}_h$.
%
%$\mathcal{E}_c \le \mathcal{E}_h$
%
Making a similar assumption as that in the HTT method, we get
$\mathcal{E}_c = \mathcal{E}_h$.
%\label{eq:PHAT_Energy_Constraint_2}
Since we assume the channel power gain process is ergodic, this equation means that as time goes to infinity, the amount of harvested and consumed energies should be equal.

\end{itemize}

Note that here we are essentially disregarding the causality constraint in a WPCN.
As time passes, however, the charge of the capacitor varies according to an unbiased one-dimensional random walk with the addition of non-negativity constraint.
Such a walk ensures that as time goes to infinity, the capacitor builds up enough energy that the probability of its depletion goes to zero.
Therefore, what we calculate should be considered as an upper bound of the throughput of such a scheme.
In a practical implementation, the decision for UL and DL should also be based on the amount of energy saved in the capacitor so as to avoid its depletion or overflow.
\textcolor{black}{For example, when the system starts up without any initial energy, the WD may only harvest energy.}

The problem described so far may be formulized as follows
\begin{subequations}
\begin{align}
& \underset{\mathcal{I},\mathcal{P}, p_u}{\text{maximize}} && \overline{r}_\mathrm{PHAT}=\int_{\mathcal{I}}\log_2 \left(1+{p_u \frac{g\bar{g}}{\sigma^2}}\right)\ f_G\left(g\right)dg\,, &&\label{pr:PHAT-GR-cost}\\ 
& \text{subject to} && \int_{\mathcal{I}} p_uf_G(g)dg = p_d \bar{g}  \int_{\mathcal{P}} g f_G(g) dg\,,  &&\label{pr:PHAT-GR-constraint1}\\ 
&&& \mathcal{I} \cap  \mathcal{P} = \varnothing\,, &&\label{pr:PHAT-GR-constraint2}\\
&&&\mathcal{I} \cup \mathcal{P} \in \mathcal{R}_+\,.&&\label{pr:PHAT-GR-constraint3}
\end{align}
\label{pr:PHAT-GR}
\end{subequations}
Observe that the sets $\mathcal{P}$ and $\mathcal{I}$ can, in general, be of any form and therefore, this problem is not convex.
Hence, we seek easier-to-analyze subsets of the permissible solutions by this formulation.

Note also that the fundamental trade-off of the WPCN system can be seen here.
Informally speaking, as set $\mathcal{P}$ includes more and more length of the whole $\mathcal{R}_+$ set, the WD harvests more and more  energy, hence being able to transmit with higher transmit power and therefore higher SNR and throughput.
Yet, this makes set $\mathcal{I}$ smaller which means the WD has less frames allocated to UL WIT.
This has the reverse effect of decreasing the throughput.

\section{Probabilistic Harvest and Transmit}
In this section, we seek solutions of optimization problem (\ref{pr:PHAT-GR}) in which WIT and WPT normalized channel power gain sets are intervals or unions of intervals.
We will name these schemes based on the relative order of these intervals on the positive real axis.
In addition, we assume the UL transmit power is constant.
This condition simplifies (\ref{pr:PHAT-GR-constraint1}) to 
\begin{equation}
p_{u,\mathrm{PHAT}} = p_d\bar{g}  \frac{\int_{\mathcal{P}} gf_G(g)dg}{\int_{\mathcal{I}} f_G(g)dg}\,.
\label{eq:UL_power-GR}
\end{equation}

As a side note, we know that the optimal UL power allocation scheme is given by water-filling \cite{Goldsmith}.
Nevertheless, we do not use this scheme because our aim is primarily to show the superiority of our duplexing method.
Obviously, using water-filling for UL information transmission leads to a definite increase in the data-rate.

\subsection{Two-interval partitioning}
\begin{figure}[t]
\includegraphics[trim=500 60 50 100,clip,scale=0.6]{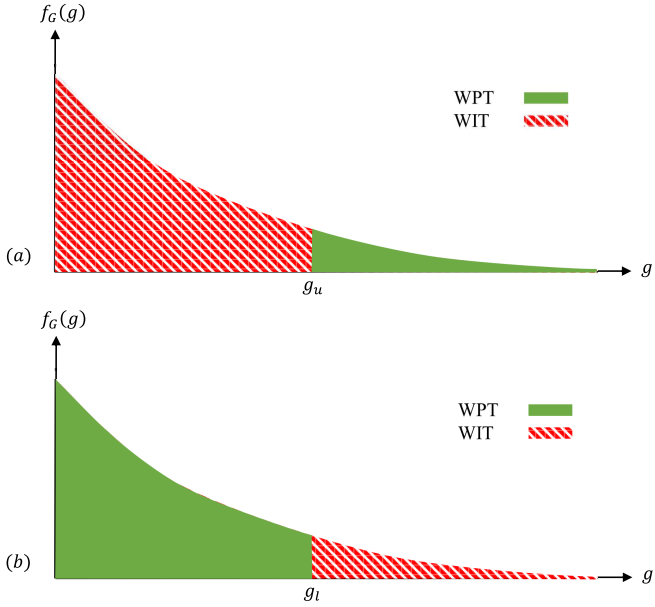}
\centering
\caption{Two-interval partitioning channel allocation scheme for a Rayleigh fading channel (a): IP (b): PI}
\label{fig:distribution_2}
\end{figure}

Here we divide the normalized channel power gain set into two intervals: the lower interval $[0~~ g)$ and the higher interval $[g~~ \infty)$.
Naturally there are two ways to assign these sets to WIT and WPT normalized channel power gain sets $\mathcal{I}$ and $\mathcal{P}$.
In Fig. \ref{fig:distribution_2} the two-interval partitioning PHAT channel allocation scheme for a Rayleigh fading distribution has been illustrated.

First, we choose the higher interval to WPT normalized channel power gain set $\mathcal{P}$, while we allocate the lower interval to WIT normalized channel power gain set $\mathcal{I}$.
Specifically, $\mathcal{I}=[0~~ g_u)$ and $\mathcal{P}=[g_u~~ \infty)$ where $g_u$ is the normalized channel power gain.
Because of the relative order of the sets $\mathcal{I}$ and $\mathcal{P}$ on the positive real axis, we call this scheme PHAT-IP.
The optimization problem (\ref{pr:PHAT-GR}) then becomes
\begin{subequations}
\begin{align}
\underset{g_u}{\text{maximize}} & &   \overline{r}_\mathrm{PHAT-IP}=\int_0^{g_u} \log_2 \left(1+\bar{\gamma}g\right)\ f_G\left(g\right)dg\,, &\label{eq:PHAT-IP-cost}\\ 
\text{subject to}               & & g_u\ge 0\,, &\label{eq:PHAT-IP-constraint}
\end{align}
\label{pr:PHAT-IP}
\end{subequations}
where $\bar{\gamma} = \frac{p_u\bar{g}}{\sigma^2}$ is the expected UL SNR.
We can now directly substitute $\mathcal{I}$, $\mathcal{P}$, and $f_G(g)$ into eq. (\ref{eq:UL_power-GR}) and find out the UL transmit power function
\begin{equation}
p_{u,\mathrm{PHAT-IP}} = p_d\bar{g}  \frac{\left(g_u+1\right)e^{-g_u}}{1-e^{-g_u}}\,.
\label{eq:UL_power-IP}
\end{equation}
In the following lemma, the ergodic throughput is calculated in closed form.

\begin{thm}
\label{lem:Data_Rate_IP}
The ergodic throughput for PHAT-IP scheme is given by
\begin{multline}
\overline{r}_\mathrm{PHAT-IP}=\frac{e^{\frac{1}{\bar{\gamma}}}}{\log 2} \left({\mathrm{E}}_1\left(\frac{1}{\bar{\gamma}}\right)-{\mathrm{E}}_{1}\left(\frac{1}{\bar{\gamma}}+g_u\right)\right)\\
-e^{-g_u}{\log_2 \left(1+\bar{\gamma}g_u\right)\ }.
\label{eq:Data_Rate_IP}
\end{multline}
where $E_1\left(z\right)$ is the exponential integral function defined as
\begin{equation*}
\mathrm{E}_1\left(z\right)=\int^{\infty }_z{\frac{e^{-t}}{t}dt}\,.
\end{equation*}
\emph{Proof}:
Substituting for $f_G\left(g\right)$ in (\ref{eq:PHAT-IP-cost}) and using integration by parts yields
\begin{equation*}
\ln 2 ~\overline{r}_\mathrm{PHAT-IP} = -e^{-g}{\left.{\ln \left(1+\bar{\gamma}g\right)\ }\right|}^{g_u}_0+\int^{g_u}_0{\frac{\bar{\gamma}}{1+\bar{\gamma}g}e^{-g}dg}\,.
\end{equation*}
But
\begin{multline*}
\int^{g_u}_0{\frac{\bar{\gamma}}{1+\bar{\gamma}g}e^{-g}dg} = e^\frac{1}{\bar{\gamma}} \int^{\frac{1}{\bar{\gamma}}+g_u}_\frac{1}{\bar{\gamma}}{\frac{1}{t}e^{-t}dt}\\
= e^\frac{1}{\bar{\gamma}} \left\{\mathrm{E}_{1}\left(\frac{1}{\bar{\gamma}}\right)-\mathrm{E}_1\left(\frac{1}{\bar{\gamma}}+g_u\right)\right\}.
\end{multline*}
Substitution gives (\ref{eq:Data_Rate_IP}).
\hfill $\blacksquare$
\end{thm}
Proving convexity of this problem is difficult.
In the following lemma, we prove the convexity of a special case of this problem.
\begin{lem}
When the HAP DL transmit power goes to infinity, optimization problem (\ref{pr:PHAT-IP}) is convex.

\emph{Proof}: 
Using the asymptotic approximations% (as $x$ approaches zero or infinity)
\begin{equation}
\lim_{\substack{x\rightarrow0}} \mathrm{E}_1(x) = \lim_{\substack{x\rightarrow\infty}} \mathrm{E}_1(x) = e^{-x}\ln\left(1+\frac{1}{x}\right)\,.
\label{eq:exponential_integral_approximation}
\end{equation}
We can find the asymptotic data rate function as $\bar{\gamma}$ tends  to infinity
\begin{equation*}
\bar{r}_\mathrm{PHAT-IP} = \log_2 \left(\bar{\gamma}\right)(1-e^{-g_u})\,.
\end{equation*}
Substituting for $\bar{\gamma}$ using (\ref{eq:UL_power-IP}) gives
\begin{equation*}
\bar{r}_\mathrm{PHAT-IP} = \log_2 \left( \frac{p_d\bar{g}^2}{\sigma^2}  \frac{\left(g_u+1\right)e^{-g_u}}{1-e^{-g_u}} \right)(1-e^{-g_u})\,.
\end{equation*}
\ifthenelse {\boolean{bombastic}}
{
\color{olive}
Taking the derivative of this equation gives
\begin{equation*}
\frac{d\bar{r}_\mathrm{PHAT-IP}}{dg_u} = \log_2 \left( \frac{p_d\bar{g}^2}{\sigma^2}  \frac{\left(g_u+1\right)e^{-g_u}}{1-e^{-g_u}} \right)e^{-g_u}-\frac{g_u+e^{-g_u}}{\ln 2(g_u+1)}\,.
\end{equation*}
\color{black}
}{}
Taking the derivative twice yields
\begin{multline*}
\frac{d^2\bar{r}_\mathrm{PHAT-IP}}{dg_u^2} = -\log_2 \left( \frac{p_d\bar{g}^2}{\sigma^2}  \frac{\left(g_u+1\right)e^{-g_u}}{1-e^{-g_u}} \right)e^{-g_u}\\
-\frac{g_u^2e^{-g_u}+2g_ue^{-2g_u}+1+3\left(e^{-2g_u}-e^{-g_u}\right)}{\ln 2\left(1-e^{-g_u}\right)\left(g_u+1\right)^2}\,.
\end{multline*}
Note that the minimum of $e^{-2g_u}-e^{-g_u}$ is $-1/4$ which is achieved at $\ln 2$.
Therefore the second derivative is negative and so the function is concave.
\hfill $\blacksquare$

\end{lem}

%The reason why our proposed scheme provides a higher throughput lies in (\ref{eq:HTT}).
%Decreasing $\tau$ decreases the harvested power and hence the data rate logarithmically, and at the same time increases the data rate linearly.
%On the other hand increasing $\tau$ increases the harvested power and hence the data rate logarithmically, and at the same time decreases the data rate linearly.
%In our scheme, however, by using the channel for DL WPT only when the normalized channel power gain is at it's peak, we are maintaining the harvested power and at the same time increasing the UL time.
%The result, is an increase in the UL data rate.
Similarly, we can allocate the higher interval to the WIT normalized channel power gain set $\mathcal{I}$ and the lower interval to the WPT normalized channel power gain set $\mathcal{P}$.
Specifically, $\mathcal{P}=[0~~ g_l)$ and $\mathcal{I}=[g_l~~ \infty)$.
\color{black}
Because of the relative order of the sets $\mathcal{P}$ and $\mathcal{I}$ on the positive real axis, we call this scheme PHAT-PI.
\color{black}
Optimization problem (\ref{pr:PHAT-GR}) then transforms to
\begin{subequations}
\begin{align}
\underset{g_l}{\text{maximize}} & &   \overline{r}_\mathrm{PHAT-PI}=\int_{g_l}^\infty \log_2 \left(1+\bar{\gamma}g\right)\ f_G\left(g\right)dg\,, &\\ 
\text{subject to}               & & g_l\ge 0\,. &
\end{align}
\label{pr:PHAT-PI}
\end{subequations}
\color{black}
Substituting $\mathcal{I}$, $\mathcal{P}$, and $f_G(g)$ into (\ref{eq:UL_power-GR}), gives the UL power
\color{black}
\begin{equation}
p_{u,\mathrm{PHAT-PI}}  = p_d\bar{g} \left( {e^{g_l}-g_l-1}\right)\,.
\label{eq:UL_power-PI}
\end{equation}
In the following theorem, \color{black}the ergodic throughput is calculated in closed form. \color{black}
The proof is similar to that of theorem \ref{lem:Data_Rate_IP} and is omitted here for the sake of brevity.
\begin{thm}
\label{lem:Data_Rate_PI}
The ergodic throughput for the PHAT-PI scheme is given by
\begin{equation}
\overline{r}_\mathrm{PHAT-PI}= e^{-g_l} \log_2 \left(1+\bar{\gamma}g_l\right) + \frac{e^{\frac{1}{\bar{\gamma}}}}{\log 2} \mathrm{E}_1\left(\frac{1}{\bar{\gamma}}+g_l\right)\,.
\label{eq:Data_Rate_PI}
\end{equation}
\ifthenelse {\boolean{bombastic}}
{
\color{olive}
\emph{Proof}:
Substituting for $f_G\left(g\right)$ in (\ref{pr:PHAT-IP}) gives
\begin{equation*}
\overline{r}_\mathrm{PHAT} \ln 2 = \int^\infty_{g_m}{{\ln \left(1+\bar{\gamma}g \right)\ }e^{-g}dg}\,.
\end{equation*}
We can use integration by parts to solve this problem
\begin{equation*}
\overline{r}_\mathrm{PHAT} \ln 2 = -e^{-g}{\left.{\ln \left(1+\bar{\gamma}g\right)\ }\right|}^\infty_{g_m}+\int^\infty_{g_m}{\frac{\bar{\gamma}}{1+\bar{\gamma}g}e^{-g}dg}\,.
\end{equation*}
But
\begin{equation*}
\int^\infty_{g_m}{\frac{\bar{\gamma}}{1+\bar{\gamma}g}e^{-g}dg} = e^\frac{1}{\bar{\gamma}} \int^\infty_{\frac{1}{\bar{\gamma}}+g_m}{\frac{1}{t}e^{-t}dt} = e^\frac{1}{\bar{\gamma}} \mathrm{E}_1\left(\frac{1}{\bar{\gamma}}+g_m\right)\,.
\end{equation*}
Substitution gives (\ref{eq:Data_Rate_PI}).
\hfill $\blacksquare$
}{}
\end{thm}
Proving convexity of optimization problem (\ref{pr:PHAT-PI}) is difficult.
In the following lemma, however, we prove that a special case of this problem is quasi-convex.
\begin{lem}
When the HAP DL transmit power goes to infinity, the optimization problem (\ref{pr:PHAT-PI}) is quasi-convex.

\emph{Proof}: 
Using the asymptotic approximation (\ref{eq:exponential_integral_approximation}), we can find the asymptotic data rate function as $\bar{\gamma}$ tends  to infinity
\begin{equation*}
\bar{r}_\mathrm{PHAT-PI} = \log_2 \left(\bar{\gamma}\right)e^{-g_l}\,.
\end{equation*}
Substituting for $\bar{\gamma}$ using (\ref{eq:UL_power-PI}) gives
\begin{equation*}
\bar{r}_\mathrm{PHAT-PI} = \log_2 \left( \frac{p_d\bar{g}}{\sigma^2} \left( {e^{g_l}-g_l-1}\right) \right)(e^{-g_l})\,,
\end{equation*}
designating the first factor as $f(g_l)= \log_2 \left( \frac{p_d\bar{g}}{\sigma^2} \left( {e^{g_l}-g_l-1}\right) \right)$
\ifthenelse {\boolean{bombastic}}
{
we can first find its first derivative
\begin{equation*}
\ln 2~ \frac{df(g_l)}{dg_l} =  \frac{e^{g_l}-1}{e^{g_l}-g_l-1}
\end{equation*}
then 
}{}
we can find its second derivative
\begin{equation*}
\ln 2~ \frac{d^2f(g_l)}{dg_l^2} =  \frac{(1-g_l)e^{g_l}-1}{\left(e^{g_l}-g_l-1 \right)^2}\,.
\end{equation*}
It is not difficult to show that this function is always negative and therefore, $f(g_l)$ is concave and as a result, log-concave.
Consequently, the ergodic throughput function is a product of log-concave and log-linear functions and is hence log-concave.
\hfill $\blacksquare$

\end{lem}

As we will see in the simulation results, the PHAT-PI scheme results in better performance in the low-SNR regime while the PHAT-IP scheme performs best in the high-SNR regime.
Consequently, in the next section we will combine these two schemes to arrive at a new scheme that has good performance in both regimes.

\subsection{Three-interval partitioning}
\begin{figure}[t]
\includegraphics[trim=60 190 510 160,clip,scale=0.6]{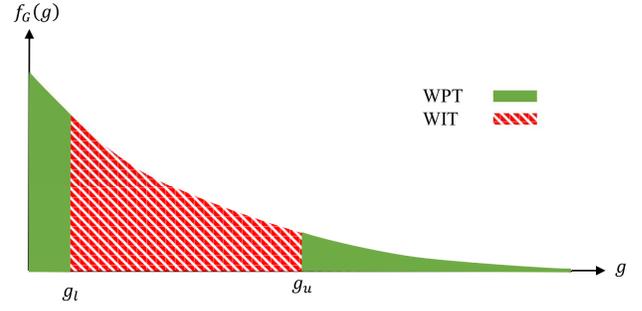}
\centering
\caption{Three-interval partitioning channel allocation scheme for a Rayleigh fading channel}
\label{fig:distribution_3}
\end{figure}

Here we divide the normalized channel power gain set into three interval: the lower interval $[0~~g_l)$, the middle interval $[g_l~~g_u)$, and the higher interval $[g_u~~\infty)$ where $g_l$ and $g_u$ are lower and upper normalized channel power gain thresholds respectively.
We let $\mathcal{P} = \mathcal{P}_1 \cup \mathcal{P}_2$, where $\mathcal{P}_1=[0~~g_l)$, and $\mathcal{P}_2=[g_u~~\infty)$; and $\mathcal{I} = [g_l~~g_u)$.
\color{black}
Because of the relative order of the sets $\mathcal{P}_1$, $\mathcal{I}$, and $\mathcal{P}_2$ on the positive real axis, we call this scheme PHAT-PIP.
In Fig. \ref{fig:distribution_3}, the three-interval partitioning PHAT channel allocation scheme for a Rayleigh fading distribution has been illustrated.

There is, obviously, another possible three-interval partitioning normalized channel power gain set allocation; that is $\mathcal{I} = \mathcal{I}_1 \cup \mathcal{I}_2$, where $\mathcal{I}_1=[0~~g_l)$, and $\mathcal{I}_2=[g_u~~\infty)$; and $\mathcal{P} = [g_l~~g_u)$.
We will not derive this scheme due to space limitation and because we found that its throughput is less than or equal to that of PHAT-PIP.

Substituting $\mathcal{I}$, $\mathcal{P}$, and $f_G(g)$ in (\ref{eq:UL_power-GR}) and calculating the integral yields
\color{black}
\begin{equation}
p_{u,\mathrm{PHAT-PIP}} = p_d\bar{g} \frac{1-(g_l+1)e^{-g_l}+(g_u+1)e^{-g_u}}{e^{-g_l}-e^{-g_u}}\,.
\label{eq:UL_power-PIP}
\end{equation}

In the following theorem, \color{black}the ergodic throughput of the PHAT-PIP scheme is calculated in closed form. \color{black}
The proof is similar to that of theorem \ref{lem:Data_Rate_IP} and is omitted here.
\begin{thm}
\label{lem:Data_Rate_PIP}
The ergodic throughput for PHAT-PIP scheme is given by
\begin{multline}
\bar{r}_\mathrm{PHAT-PIP} = e^{-g_l}\ln\left(1+\bar{\gamma}g_l\right) - e^{-g_u}\ln\left(1+\bar{\gamma}g_u\right)\\
 + e^\frac{1}{\bar{\gamma}} \left(\mathrm{E}_1\left(\frac{1}{\bar{\gamma}}+g_l\right)-\mathrm{E}_1\left(\frac{1}{\bar{\gamma}}+g_u\right)\right)\,.
\label{eq:Data_Rate_PIP}
\end{multline}
\end{thm}

Note that neither this function, nor its infinite-SNR asymptotic approximation are (log)-concave.
As a result we can only use  exhaustive search to find the optimal value of this problem.

\begin{figure}[t]
\includegraphics[trim=260 60 300 130,clip,scale=0.6]{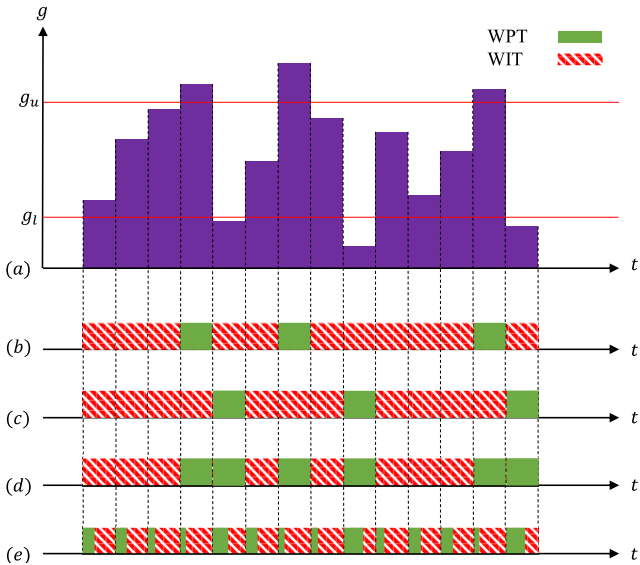}
\centering
\caption{Comparison of the HTT, and PHAT schemes. (a) normalized channel power gain random process, (b) the PHAT-IP scheme, (c) the PHAT-PI scheme, (d) the PHAT-PIP scheme, (e) the HTT scheme }
\label{fig:time-series}
\end{figure}

In Fig. \ref{fig:time-series} the three channel allocation schemes have been illustrated and compared with the HTT protocol.
As can be seen, by setting $g_l=0$ this scheme reduces to PHAT-IP and by letting $g_u\rightarrow\infty$ it reduces to PHAT-PI.
\color{black}
%Observe that in the PHAT schemes, the system may not be in the UL mode for a few consecutive frames, whereas, in the HTT scheme, a portion of every frame is necessarily allocated to the UL in every frame.
%This means that these schemes incur higher latency.
%Note that, the complexity of the PHAT-PI and PHAT-IP schemes are the same as that of HTT.
\color{black}

\section{Simulatoin Results}
\ifthenelse {\boolean{bombastic}}
{
\color{olive}
\subsection{Setup}
\color{black}
}{}
\color{black}
In this section, the performances of the PHAT schemes are evaluated and compared against the HTT scheme.
%The performance is measured based on the SNR value and is therefore independent of the complete simulation setup.
For the HTT scheme, 100000 channel realizations are used to numerically evaluate integral (\ref{eq:Ergodic}).
In order to search the space in the PHAT schemes, a somewhat arbitrary value of 10 was assumed for the maximum achievable channel power gain.
Thanks to their (quasi) convexity, the optimization problem of HTT, PHAT-PI and PHAT-IP schemes were solved using the bisection method.
On the other hand, the PHAT-PIP method was solved using 2-D exhaustive search.
%Note that the PHAT-PIP scheme, requires 2-D exhaustive search which is much more expensive than 1-D bisection.
%Yet, in a static environment where the fading distribution does not change with time, the calculation, once done, does not need to be repeated for every frame which is the case for HTT.

\color{black}

\begin{figure}[t]
\includegraphics[trim = 40 10 40 10,clip,scale=0.48]{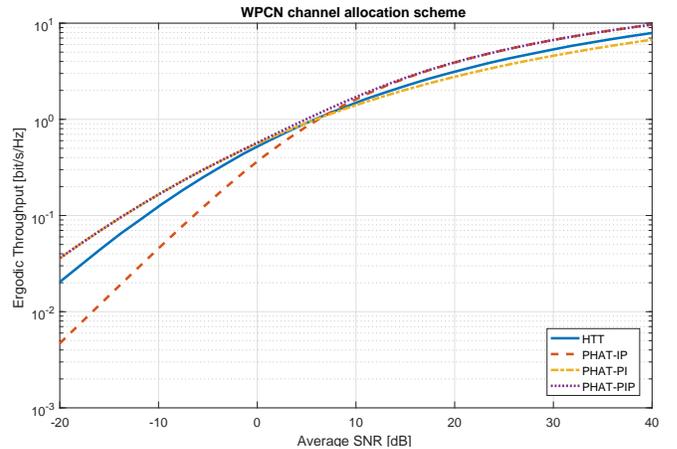}
\centering
\caption{Throughput of the system versus average SNR for HTT and PHAT schemes}
\label{fig:comparison1}
\end{figure}

\ifthenelse {\boolean{bombastic}}
{
\color{olive}
\subsection{Results}
\color{black}
}{}
As can be seen in Fig. \ref{fig:comparison1}, at high enough SNR values, the throughput of the PHAT-IP scheme has exceeded that of the HTT scheme.
Nevertheless, at low SNR values, the HTT scheme can outperform the PHAT-IP method.
Conversely, at low SNR values, the throughput of the PHAT-PI scheme is higher than that of the HTT scheme.
Yet, three-interval partitioning of the normalized channel power gain distribution has resulted in an increase in the throughput both in the low-SNR as well as the high-SNR regimes.
In the low-SNR region, the power is mainly harvested from the higher interval of the distribution, whereas in the high-SNR region, the power is mainly harvested from the lower interval of the distribution.
%This means that in a three-interval-partitioning setting, the upper and lower thresholds both decrease as the SNR increases.
%

\section{Conclusion}
In this paper, we showed that rather than dividing every frame to WIT and WPT phases, careful allocation of the whole frames to WIT and WPT results in a definite increase in the throughput of a WPCN.
Such an allocation is based on the value of the estimated normalized channel power gain.
The implementation of such a scheme, however, involves an infinite capacitor, and the throughput is reached at time infinity.
A practical design taking account of the causality constraint and using a capacitor of finite capacity can be envisioned.
We conceive that in such a scheme the thresholds, in addition to the estimate of the current channel power gain, will be dependent on the current charge of the capacitor.
Such a design can be the topic of a future work.

\ifthenelse {\boolean{bombastic}}
{
\newpage
\color{olive}
\section{Notes}

\begin{itemize}
\subsection{Color convention}
\item Olive: This text belongs to the extended version (For my own reference)
\item Red: Checkpoints to navigate through the paper more easily
\item Green: Might be incorrect! Please check!
\item Blue: Have just added this... needs immediate revision

\subsection{Things to be added}
\item The high-SNR asymptotic was shown to be convex. Explain why bisection is used.
\item Checking figure locations and where they are mentioned
\item Emphasizing our contribution
\item Printing the name of all labels

\subsection{Things to investigate}
\item Changing the sentences to active voice
\item Check which equation should be in display mode
\item Check Vocabulary
\item Check ``THE''
\item Check Initial energy (?)
\item Double antennas at the WD?
\item Paragraph continuity

\end{itemize}

}{}

\end{document}